\documentclass[12pt]{iopart}
\usepackage{xcolor}
\usepackage{iopams}
 \usepackage{amssymb}
\usepackage{mathrsfs}
\usepackage[mathscr]{euscript}
\usepackage{slashed}

\def\sgn{\mathrm{sgn}}
\begin{document}

\title[Fractional exclusion and braid statistics in one dimension]{Fractional exclusion and braid statistics in one dimension: a
  study via dimensional reduction of Chern-Simons
  theory\footnote{Dedicated to the memory of Mario Tonin.}}

\author{Fei Ye$^1$, P. A. Marchetti$^2$, Z. B. Su$^3$ and L. Yu$^{4,5}$} 
\address{$^1$ Department of Physics, South University of Science and
   Technology of China, Shenzhen 518055, China}
\ead{yef@sustc.edu.cn}

\address{$^2$ Dipartimento di Fisica, INFN, I-35131 Padova, Italy}
\ead{marchetti@pd.infn.it}

\address{$^{3}$ Institute of Theoretical Physics, Chinese Academy of
  Sciences, 100190 Beijing, China}

\address{$^{4}$ Institute of Physics, Chinese Academy of Sciences, and Beijing National Laboratory for Condensed Matter Physics, 100190
 Beijing, China}
\address{$^{5}$ Collaborative Innovation Center of Quantum Matter, 100190 Beijing, China}

\begin{abstract}
  The relation between braid and exclusion statistics is examined in
  one-dimensional systems, within the framework of Chern-Simons
  statistical transmutation in gauge invariant form with an appropriate
  dimensional reduction.  If the matter action is anomalous, as for
  chiral fermions, a relation between braid and exclusion statistics can
  be established explicitly for both mutual and nonmutual
  cases. However, if it is not anomalous, the exclusion statistics of
  emergent low energy excitations is not necessarily connected to the
  braid statistics of the physical charged fields of the
  system. Finally, we also discuss the bosonization of one-dimensional
  anyonic systems through T-duality.
\end{abstract}
\maketitle

\section{Introduction}
\label{sec:introduction}
In quantum mechanics of identical particles, there are two ways, a priori
rather different, to define the statistics.

One is the \emph{braid}(or exchange) statistics, which can be defined
through the monodromy of the many-body wave-function as follows: For
$\mathbb{C}$-valued wave functions, when one performs a positively
oriented exchange between two particles, the wave function acquires a
phase factor $e^{i(1-\alpha)\pi}$, with $\alpha$ as the braid-statistics
parameter and in space-dimensions $d<3$. It is in general arbitrary. In
fact considering all possible oriented exchanges in an $n$-particle wave
function these phase factors provide an abelian representation of the
braid group $B_n$\cite{Wu1984a,Frohlich1988a}. Obviously,
fermions(bosons) correspond to $\alpha=0(\alpha=1)$; for
$\alpha \neq 0,1$ the statistics is called fractional. In one or two
dimensional space, the existence of \emph{braid} statistics is well
known \cite{Leinaas1977,Goldin1981,Wilczek1982}, and the corresponding
particle is dubbed as \emph{anyon}. The first example of
  what will be called a braid statistics of fields in $d=1$ appears for
  the free fields in Ref.~\cite{Streater1970}. The earliest examples in
  interacting theories can be found in nuce in Ref.~\cite{Frohlich1976},
  and are discussed in details in Ref.~\cite{Fradkin1980} and in
  Ref.~\cite{Frohlich1992a}.  According to Wilczek\cite{Wilczek1982},
the anyon in 2D can be viewed as a charged particle(fermion or boson)
bound to a flux with the statistics coded through the Aharonov-Bohm
effect. The flux-binding changing the braid statistics can be realized
through a minimal coupling to a Chern-Simons gauge field and this
procedure is called statistical transmutation\cite{Polyakov1988}. For
$d\ge3$ only $\alpha =0$ or $1$ are allowed, since the orientation of
the exchange is irrelevant and the braid group collapses to the
permutation group.

The other kind of statistics can be defined through state counting,
which is known as (Haldane's) \emph{exclusion} statistics and can be
viewed as an effective interaction among particles occupying identical
or different states in the Hilbert space. The exclusion statistics is
characterized by a parameter $g$, first introduced by Haldane
\cite{Haldane1991a}, which measures the change rate of the dimension $D$
of Hilbert space with respect to the total particle number $N$ when an
additional particle is introduced. For a single species of particles,
there is a linear relation between $D$ and $N$,
$\Delta D=(g-1)\Delta N$. Again, $g=0$ for fermions and $g=1$ for
bosons. It turns out that $1/(1-g)$ gives also the maximum average
occupation number for the quantum states below the Fermi
energy\cite{Wu1994}. The fractional \emph{exclusion} statistics can be
viewed as a generalized Pauli's exclusion principle and it exists in
arbitrary dimensions\cite{Haldane1991a}.  Following the nomenclature in
Refs.\cite{WuYu1995}, we call \emph{exclusons} the low-energy
quasiparticles/quasiholes obeying fractional exclusion statistics.  The
best known examples of exclusons arise in a number of one dimensional
systems solvable by the thermodynamic Bethe ansatz, such as the
Yang-Yang $\delta$-function gas \cite{Lieb1963a,Lieb1963b,yang1969} or
the Calogero-Sutherland model
\cite{Calogero1969b,Sutherland1971a,Sutherland1971b,Sutherland1971c,Sutherland1972}.
In two dimensions there are also a few strongly correlated systems whose
low-lying excitations exhibit the fractional exclusion statistics,
\cite{Haldane1991a,Wu1994,deVeigy1994}, including the "Laughlin
vortices" of the fractional quantum Hall effect.  For relativistic
elementary particles the fermion(boson) exclusion statistics can be
derived from the antisymmetry(symmetry) of the many-body wave-function
and therefore the two ways of defining the statistics can be identified.

Both types of statistics are naturally related to interactions. In the
Chern-Simons theory of anyons, the braid statistics arises simply as a
charge-current interaction among particles mediated by the statistical
gauge fields (see \cite{Ye2015}). Although the anyon model is easy to
construct using Chern-Simons statistical transmutation, even the 2D free
anyon gas is extremely difficult to solve due to the highly entangled
motions of anyons. The exclusion statistics is also a consequence of
interactions. It is actually an emergent phenomenon in the low-energy
behavior of an interacting system consisting of physical particles with
prescribed braid statistics(usually either fermion or boson).  The
exclusion statistics can be directly applied to calculate the
thermodynamics of the excluson gases\cite{Wu1994}. As a typical example,
the Yang and Yang's thermodynamic Bethe ansatz solution of the one
dimensional repulsive $\delta$-potential boson gas can be reformulated
as a \emph{free} excluson gas obeying nontrivial mutual exclusion
statistics\cite{Bernard1994anote}.  More generally, in one dimensional
integrable models the thermodynamic Bethe ansatz equations can be
reinterpreted exactly as statistical interactions between exclusons with
identical or different momenta, determined by the two-body scattering
phase shift\cite{Bernard1994anote}. Based on these observations, Wu and
Yu proved that the low energy physics of an excluson gas is equivalent
to that of a one-component Luttinger liquid with Haldane's controlling
parameter $\lambda$ identified as the statistical parameter
$1-g$\cite{WuYu1995,Wu2001}, thus providing a unified description for
various interactions using fractional exclusion statistics.  Inspired by
the success in one dimensional systems, there were attempts to
generalize the Fermi liquid theory to a ``Haldane'' liquid with
fractional exclusion statistics in higher dimension to provide a unified
description to the low energy behavior of interacting
systems\cite{Iguchi1998}.  The fractional statistics(both exchange and
exclusion) has also been proposed to provide a better mean field theory
for some strongly correlated systems, especially the cuprate
superconductors\cite{Laughlin1988,Frohlich1992,Marchetti1998,Marchetti2011a}.

A natural question to ask is what is the relation between the two
aspects of the fractional statistics, i.e. between $\alpha$ and $g$. It
turns out that there is no  universal relationship, but relations appear
in specific examples. In two dimensional quantum Hall systems with Hall
conductance $\sigma_h$, we can derive a linear relation between $g$ and
$\alpha$, $g=2\pi\sigma_h\alpha$\cite{Ye2015}, clarifying and extending
previous results. More generally we proved the existence of a relation
in sytems with chiral edge currents.  Although there were many studies
on the fractional statistics in one dimension, to the best of our
knowledge, it is still lacking a satisfactory understandings of the
relation between $g$ and $\alpha$, probably in part because the exchange
of two particles in one dimension necessarily involves scattering
processes, thus there is no unique way to separate the braid statistics
from the dynamical
processes\cite{Bernard1994anote,Ha1994b,Ha1995,WuYu1995,Wu2001}.

In this article, we examine the relation between braid and exclusion
fractional statistics of particles moving in a straight line, including
chiral cases.  In the present study, we adopt the fermion-based
Chern-Simons theory to unambiguously define the braid statistics. This
is also the same framework adopted in our previous study on two
dimensional fractional statistics\cite{Ye2015}, but to discuss braid
statistics in one dimension we need a careful dimensional reduction
following the technique developed in \cite{Marchetti1996}. In this
setup, we restrict to impenetrable two-body interactions, to have a
well-defined braid statistics. The main purpose of this article is then
to study \emph{whether the Chern-Simons statistical transmutation can
  induce non-trivial fractional exclusion statistics or not}.  It turns
out that if the gauge effective action of the matter system minimally
coupled to the statistical gauge field is gauge invariant, the exclusion
statistics of the emergent exclusons has nothing to do with the braid
statistics of the physical charged fields of the model, but the
statistical transmutation shifts the value of their Fermi momenta.
However, if the gauge effective action of the matter system is
anomalous(chiral case), then a precise relation between the braid and
exclusion statistics emerges.  Thus, our present study for
one-dimensional systems together with our previous study for
two-dimensional systems provides a systematic description of the
relation between fractional abelian exchange statistics and fractional
exclusion statistics in low dimensions($d<3$) in the same framework of
Chern-Simons theory.

This article is organized as follows: in
Sec.~\ref{sec:chern-simons-theory} the dimensional reduction of two
dimensional Chern-Simons theory is introduced, and with this formalism
we calculate the Green's functions of noninteracting anyons. In
Sec.~\ref{sec:fract-braid-excl} we present the results in the presence of
both fractional exchange and exclusion statistics for matter systems
that exhibit a gauge-invariant effective action if minimally coupled to
the Chern-Simons gauge field.  In Secs.~\ref{sec:nonm-stat-single-1} and
\ref{sec:mutu-stat-mult}, we analyze the relation between the braid and
exclusion statistics emerging for \emph{noninteracting} anyonic systems
for both nonmutual and mutual statistics.  Finally, in
Sec.~\ref{sec:bosonization} we sketch a derivation of the corresponding
bosonization formulas for one-dimensional anyonic system via
T-duality. Throughout the paper we use the euclidean path-integral
formalism.

\section{Chern-Simons theory and dimensional reduction for
  noninteracting Anyons}
\label{sec:chern-simons-theory}
We consider non-relativistic spinless fermions in one dimension with
particle density $\rho^0$ and Fermi velocity $v_f$ set to unit for
simplicity. In the scaling limit, the low energy physics is controlled
by excitations near the two Fermi points $\pm k_f$ with
$k_f = \pi \rho^0$ in the noninteracting case. This allows us to
decompose the non-relativistic fermion field $\Psi$ in the low-energy
region into right and left movers as
$\Psi(x) \sim \psi_R(x)e^{ik_fx^1} + \psi_L(x) e^{-ik_fx^1}$, with the
dynamics of a massless Dirac fermion described by the spinor doublet
$\psi=(\psi_R,\psi_L)^t$. For convenience we also use $\psi_{1,2}$ to
denote $\psi_{R,L}$, respectively.

To implement the braid statistics, we first embed the 1+1 dimensional
spacetime of the matter field with coordinates $x^0$ and $x^{1}$, into
the 2+1 dimensional spacetime as the plane at $x^2=0$.  Then we couple
minimally the matter fields to a 2+1 dimensional statistical gauge field
$A_{\mu}$, whose dynamics is described by the Chern-Simons action. The
total action in Euclidean space with metric tensor
$g_{\mu\nu}=\mathrm{diag}(-1,-1,-1)$ consists of the following two terms:
\begin{eqnarray}
\label{eq:1}
   S_f[\psi, \bar\psi \vert A]=
    \int d^2x \bar{\psi} i\gamma^{\mu}D_{\mu}\psi
-i\int d^2x \rho^0 A_0
, \\
\label{eq:2}
  S_{cs}[A]= \frac{i}{4\pi \alpha} \int d^3x \epsilon^{\mu\nu\lambda}A_{\mu}\partial_{\nu}A_{\lambda}, 
\end{eqnarray}
where $\rho^0$ is the expectation value of the fermion density with
respect to the noninteracting vacuum, and the covariant derivatives are
defined as $D_{\mu} = \partial_{\mu}+i A_{\mu}(x, x^2=0)$ with
$x=(x^0,x^1)$ and $\mu=0,1$.  The $\gamma$-matrices 
$\{\gamma^{\mu},\gamma^{\nu}\}=2 g^{\mu\nu}$, and
$\bar{\psi}=-i\psi^{\dagger}\gamma^0$.

Being the matter field one-dimensional, one needs to carry out the
dimensional reduction from three to two dimensions, implemented
following the technique developed in Ref.\cite{Marchetti1996}. Since the
$A_0$ field has no dynamics, it can be integrated out leading to the
following flux-binding constraint,
\begin{eqnarray}
\label{eq:3}
F_{12}(x,x_2)=-2\pi i \alpha\delta(x^2)[j^0(x)+i\rho^0],
\end{eqnarray}
where $j^\mu=:\bar{\psi}\gamma^\mu \psi:$ is the (normal ordered) Dirac current.
\eref{eq:3} has the following solution:
\begin{eqnarray}
\label{eq:4}
A_1(x,x^2) = i\pi\alpha\sgn(x^2)[j^0(x)+i\rho^0]+\partial_1f(x,x^2), \nonumber\\
A_2(x,x^2)=\partial_2f(x,x^2), 
\end{eqnarray}
where $f(x,x_2)$ is an arbitrary gauge function and the sign function is
taken antisymmetric, i.e., $\sgn(0)=0$. Inserting Eq.~\eref{eq:4} into
the remaining term of the Chern-Simons action and in Eq.~\eref{eq:1},
we obtain a free Dirac fermion coupled to a pure gauge field
$\partial_{\mu}f$, and by choosing the gauge $A_2=0$ the gauge function
$f$ can be absorbed by a redefinition of the fermion field, as we assume
henceforth. It follows that the energy spectrum and the correlators of
the Fermi field $\psi$ are controlled by local \emph{fermionic}
excitations, unaffected by flux binding. This is due to the triviality
of the total partition function of the gauge field, leaving only the
free fermion as the final result.

However, the correlators of the $\psi$ field alone are not gauge
invariant, hence they are unphysical. Therefore, the states obtained
from the vacuum acting with the Fermi field $\psi$ do not belong to the
physical Hilbert space of the theory with action given by
Eqs.~\eref{eq:1} and \eref{eq:2}. In fact it is known that in gauge
theories the physical charged excitations can be created/annihilated by
gauge invariant non-local fields acting on the physical Hilbert
space(see Ref.\cite{Marchetti2010} and references therein). In the
present case for $\alpha \neq 0,1$ these fields obey non-trivial braid
statistics and can be constructed as follows: The action $S_f$ is
invariant under gauge transformation $\psi\rightarrow e^{-i\Lambda}\psi$
and $A_{\mu}\rightarrow A_{\mu}+\partial_{\mu}\Lambda$; correspondingly
the gauge-invariant anyon field is given by:
\begin{eqnarray}
\label{eq:5}
\tilde{\psi}_{R,L}(x)\equiv\psi_{R,L}(x) e^{-i\int_{P_x}A_{\mu}d\ell^{\mu}},
\end{eqnarray}
where the path $P_x$ is a straight line from $x^1$ to $\infty$ with
fixed time $x^0$.  To avoid an ill-defined crossing with the world lines
of the fermions, we shift $P_x$ slightly from $x^2=0$ to $x^2=\epsilon$
with $\epsilon$ an infinitesimal positive number. More precisely we take
the limit $\epsilon \searrow 0$ on the correlation functions and
technically one needs a compensating current at infinity joining all the
paths $P$ appearing in the correlators, but we do not discuss this
matter here, referring for details to
Refs.\cite{Frohlich1992,Marchetti1996}. The exponential in the r.h.s. of
Eq.~\eref{eq:5} will be called gauge string. The low-energy anyon
Green's function including the Fermi momenta is given by:
\begin{eqnarray}
\label{eq:6}
  G^{\alpha}_{ab}(x,y) = \langle \tilde{\psi}_a(x)
  \tilde{\psi}_b^{\dagger}(y)\rangle e^{-i\pi \rho^0[(-1)^a x^1-(-1)^b y^1]}
\end{eqnarray}
with $a = 1,2$ and the same for $b$, where the expectation
value is now referred to zero density.  With the above prescription the
braiding effect is captured by the gauge strings.

To calculate the Green's function of the physical particles
$G^{\alpha}_{ab}(x,y)$, one needs to average Eq.~\eref{eq:6} over the
statistical gauge field $A_{\mu}$ weighted by the Chern-Simons action,
following Ref.\cite{Marchetti1996}.  We first insert Eq.~\eref{eq:4}
into Eq.~\eref{eq:5}, yielding
\begin{eqnarray}
\label{eq:7}
  \tilde{\psi}_a(x)=\psi_a(x)e^{-i\pi\alpha\rho^0x^1+\pi\alpha\int_{x^1}^{\infty}j^0(x^0,z^1)dz^1}.
\end{eqnarray}
Then the Green's function defined by Eq.~\eref{eq:6} reads
\begin{eqnarray}
\label{eq:8}
 G^{\alpha}_{ab}(x-y)  
&=\frac{1}{Z_{0f}}
              e^{-i\pi\rho^0[(-1)^{a}x^{1}-(-1)^{b}y^{1}+\alpha(x^1-y^1)]} \nonumber\\
 &\times
 \int \mathcal{D}\psi \mathcal{D}\bar{\psi}\psi_a(x)\psi^{\dagger}_b(y) 
e^{\int d^2x\bar{\psi}i\gamma^{\mu}[\partial_{\mu}-i\pi\alpha\delta_{\mu}^0\partial_0Q_{x,y}(z)]\psi}
\end{eqnarray}
where
$Z_{0f}\equiv{\int \mathcal{D}[\psi,\bar{\psi}]
  e^{S_f[\psi,\bar{\psi}|0]}}$ is the partition function of free Fermi
field and
\begin{eqnarray}
\label{eq:9}
Q_{x,y}(z) 
\equiv&\theta(z^1-x^1)\theta(z^0-x^0) 
-\theta(z^1-y^1) \theta(z^0-y^0)
\end{eqnarray}
with $\theta(x)$ being the step function. To calculate the Green's
function Eq.~\eref{eq:8}, we follow Schwinger's
formalism\cite{Schwinger1962}, which leads to the exact correlation of
the Fermi fields in the presence of gauge field,
\begin{eqnarray}
\label{eq:10}
G_{ab}(x-y|\mathcal{A})&\equiv \frac{1}{Z_{0f}} \int \mathcal{D}\psi
  \mathcal{D}\bar{\psi} \psi_a(x)\psi_b^{\dagger}(y)e^{\int
  d^2z\bar{\psi}i\gamma^{\mu}(\partial_{\mu}+i\mathcal{A}_{\mu})\psi}  \nonumber\\
&= \frac{\delta_{ab}}{2\pi} 
\frac{e^{i\Theta^a_{x,y}[\mathcal{A}]}
e^{S_{eff}[\mathcal{A}]}}{(x^0-y^0)+i(-1)^a(x^1-y^1)},
\end{eqnarray}
where the functionals $\Theta^a_{x,y}$ and $S_{eff}$ are given by
\begin{eqnarray}
\label{eq:11}
\Theta^a_{x,y}[\mathcal{A}_{\mu}]=&\int d^2z[\Delta^{-1}(z-x)-\Delta^{-1}(z-y)]
                          \nonumber\\ 
&\times[\partial^{\mu}\mathcal{A}_{\mu}(z)-i(-1)^a\epsilon_{\mu\nu}\partial^{\mu}\mathcal{A}^{\nu}(z)], 
  \\ 
\label{eq:12}
S_{eff}[\mathcal{A}_{\mu}] =& \frac{1}{2\pi}\int d^2z
\epsilon^{\mu\nu}\partial_{\mu}\mathcal{A}_{\nu}(z)\Delta^{-1}
\epsilon^{\sigma\tau}\partial_{\sigma}\mathcal{A}_{\tau}(z),
\end{eqnarray}
with $\Delta\equiv-\partial_{\mu}\partial^{\mu}$ being the two
dimensional Laplacian.  Comparing Eq.~\eref{eq:8} with the Schwinger's
formula Eq.~\eref{eq:10}, by identifying
$\mathcal{A}_{\mu}=-\pi\alpha\delta_{\mu}^0\partial_0Q_{x,y}$, one
immediately finds the gauge-invariant anyon's Green's function
\begin{eqnarray}
\label{eq:13}
G^{\alpha}_{ab}(x-y)=\frac{\delta_{ab}}{2\pi}
\frac{e^{-i{[}\alpha+(-1)^{a}{]}{[}\pi\rho^0(x^1-y^1)+\arg(x-y){]}}}{|x-y|^{1+(-1)^a\alpha+\alpha^2/2}}.
\end{eqnarray}
where we denote by $\arg(x)$ the argument of the complex number
$x^0+i x^1$. 

The Green's function Eq.~\eref{eq:13} indicates that there is no
correlation between the left- and right-handed branches. When $\alpha=0$
one easily recovers the free fermion result, and when $\alpha=1$, the
Green's function of the right-handed branch has the well-known form
$|x-y|^{-1/2}$ of the one-dimensional hardcore bosons\cite{Gogolin1998}.
The braid statistics of the physical particles is reflected in the
numerator of Eq.~\eref{eq:13}. If we exchange $x$ and $y$ with
increasing the argument of $x-y$ by $\pi$, an additional phase
$e^{-i\alpha\pi}$ appears besides the Fermi statistical factor.

The free fermions obey the the Pauli's exclusion principle, and each
particle occupies a volume of $2\pi/L$ for a finite system of length $L$
in the (pseudo-) momentum space. This gives rise to a finite Fermi area
determined by the fermion density $\mathcal{S}=2\pi \rho$. For the
general exclusons with non-mutual statistical interaction
$\lambda\equiv 1-g$, the occupied volume per particle is modified to
$2\pi \lambda/L$, then the ``Fermi area'' is also changed with
$\delta\mathcal{S}/(\delta\rho)=2\pi\lambda$.  Since in our calculation
the particle density is kept invariant when the interactions are
switched on, the exclusion statistics is eventually reflected in the
change of the Fermi area.  We read the Fermi momenta off from the
coefficient of $(x^1-y^1)$ in the phase factor. Indeed, the Fermi points
are shifted by the Chern-Simons coupling from $\pm \pi\rho^0$ to
$(\pm 1-\alpha)\pi\rho^0$. However, the Fermi area is still
$2 \pi\rho^0$. Hence there is no sign of non-trivial exclusion
statistics.

\section{Fractional braid and exclusion statistics in anyon "Luttinger"
  Liquids}
\label{sec:fract-braid-excl}
In the previous section we have considered the one dimensional free
anyon systems, where the elementary excitations still obey the
conventional Pauli exclusion principle in spite of their anyonic nature.
Once the two-body interaction is turned on in one dimensional fermion
gases, the system may become in the low-energy limit an \emph{excluson}
gas (we adopt the jargon invented in Refs.\cite{WuYu1995}) subject to
the fractional exclusion
statistics\cite{Haldane1991a,Bernard1994anote,WuYu1995}, for which the
low-energy physics can be described by a Luttinger liquid theory
\cite{WuYu1995,Wu2001}, as quoted in the introduction. Since the
Luttinger liquid theory is applicable to a wide class of one-dimensional
systems, one may expect that the fractional exclusion statistics is
ubiquitous in one dimension.

In fact the exclusion statistics can also be introduced in the anyon
systems. By minimally coupling a one-dimensional Luttinger liquid to a
Chern-Simon gauge field, the two types of statistics can be realized
simutaneously. A standard way to introduce a non-trivial exclusion
statistics is to add to the action Eq.~\eref{eq:1} an interaction term
in the Luttinger-Thirring form $(\kappa \pi/2) j_{\mu}(x)j^{\mu}(x)$,
which one can rewrite (up to a UV renormalization) by introducing a
vector Hubbard-Stratonovich(H.S.) field $B_\mu$ as
\begin{eqnarray}
\label{eq:14}
  \int d^2 x 
  \left[\frac{1}{2 \kappa \pi} B_\mu B^{\mu} (x) -  B_\mu j^{\mu}(x)\right].
\end{eqnarray}
As result, the total partition function has the following form
\begin{eqnarray}
\label{eq:15}
Z_T=\int \mathcal{D}A \mathcal{D}B \mathcal{D}\psi \mathcal{D}\bar{\psi} 
     e^{S_f[\psi,\bar{\psi}|A+B]+S_{cs}[A]+\frac{1}{2\pi \kappa}\int d^2zB_{\mu}B^{\mu}},
\end{eqnarray}
and the Green's function of the anyon fields reads
\begin{eqnarray}
\label{eq:16}
   &G_{ab}^{\alpha,\kappa}(x-y)  \nonumber\\
  =& \frac{1}{Z_T}e^{-i\pi\rho^0[(-1)^ax^1-(-1)^by^1]} \nonumber\\
& \times     \int \mathcal{D}A \mathcal{D}B \mathcal{D}\psi
     \mathcal{D}\bar{\psi} \tilde{\psi}_a(x)\tilde{\psi}^{\dagger}_b(y)
     e^{S_f[\psi,\bar{\psi}|A+B]+S_{cs}[A]+\frac{1}{2\pi \kappa}\int d^2zB_{\mu}B^{\mu}}.
\end{eqnarray}

The procedure is then similar to that given in
Sec.~\ref{sec:chern-simons-theory}, except that an additional procedure
of integrating over the H.S. auxiliary field $B$ is needed now. We first
integrate over $A_0$ leading to the same constraint, Eq.~\eref{eq:4},
which is then substituted into Eq.~\eref{eq:1} and
Eq.~\eref{eq:2}. Now we obtain
\begin{eqnarray}
\label{eq:17}
& G^{\alpha,\kappa}_{ab}(x-y) \nonumber\\
=&
\frac{\delta_{ab}}{2\pi}
\frac{e^{-i\pi\rho^0(-1)^{a}(x^1-y^1)}}{(x^0-y^0)+i(-1)^a(x^1-y^1)} \nonumber\\
 &\times
\frac{\int \mathcal{D}Be^{\int
  d^2z \frac{B_{\mu}B^{\mu}}{2\pi\kappa}+i\Theta^a_{x,y}[B+\mathcal{A}]+S_{eff}[B+\mathcal{A}]-i\int
  d^2z \rho^0(B_0+\mathcal{A}_0)}}
{\int \mathcal{D}Be^{\int
  d^2z\frac{B_{\mu}B^{\mu}}{2\pi\kappa}+S_{eff}[B]-i\int d^2z\rho^0B_0}},
\end{eqnarray}
where $\mathcal{A}_{\mu}=-\pi\alpha\delta_{\mu}^0\partial_0Q_{x,y}(z)$
as before. Using Eq.~\eref{eq:11} and Eq.~\eref{eq:12}, integrating
over $B$-field is straightforward, though a little bit tedious, and it
leads to the following Green's function in the presence of both
Chern-Simons term and Thirring interaction:
\begin{eqnarray}
\label{eq:18}
 G^{\alpha,\kappa}_{ab}(x-y)=\frac{\delta_{ab}}{2\pi} e^{-i[\alpha+(-1)^a]
           [\frac{\pi\rho^0(x^1-y^1)}{1+\kappa}+\arg(x-y)]} 
 |x-y|^{-\frac{(1+\kappa)^2+[\alpha+(-1)^a]^2}{2(1+\kappa)}}. 
\end{eqnarray}

The Green's function Eq.~\eref{eq:18} goes back to the standard
power-law form of the Luttinger liquid theory, when $\alpha=0$.  The
Haldane's controlling parameter can be read off from the decaying
exponent $\lambda=1/(1+\kappa)$, which is also the exclusion parameter
according to Y.S. Wu and Y.Yu\cite{WuYu1995}. One can also identify the
exclusion statistics by fixing the particle number and directly
measuring the occupied area in the pseudo-momentum space after the
interactions are switched on. As explained in the previous section, the
change of the Fermi area is in fact a direct consequence of the
nontrivial exclusion statistics. One can read the left and right Fermi
wavevectors from the Green's function which are
$\pm\pi\rho^0/(1+\kappa)$, respectively.  The corresponding Fermi area
is then $2\pi\rho^0/(1+\kappa)$ indicating a statistical interaction
$\lambda=1/(1+\kappa)$.  This Fermi area is not changed even when the
corresponding fields becomes anyonic with braid parameter $\alpha$,
implying that the braiding effect is not necessarily connected to the
exclusion statistics. This conclusion is in fact quite general, since it
is obtained in the framework of one-dimensional interacting Dirac
fermions coupled to the Chern-Simons gauge field and the Dirac Fermion
describes the low energy physics of a large class of one dimensional
models. Our result is also consistent with that given in
Ref.\cite{Aglietti1996}.

Before ending this section, we discuss the periodicity of $\alpha$ from
the point of view of Chern-Simons theory.  As well known, for a finite
number of non-relativistic particles in the first-quantization
formalism, there is a period $2$ for the braid parameter $\alpha$, since
binding $4\pi$-flux does not change the exchange statistics. However, we
notice that in the low-energy description in the thermodynamic limit,
the shift $\alpha\rightarrow\alpha+2n(n\in\mathbb{Z})$ is not trivial
due to the coupling to the average particle density, which in fact
corresponds to multiple particle-hole excitations between the two Fermi
points with current
$2nk_f$($\kappa_{f}=\pi\rho^0$)\cite{Haldane1981a}. In general cases
with more sophisticated dispersion and interaction, the generic form of
the Green's function for the physical anyons obeying the same braid
statistics $1-\alpha$ is actually a sum of $G^{\alpha+2n}_{RR}$,
$n\in \mathbb{Z}$, with $\alpha$ restricted in the range $[0,2)$,
\begin{eqnarray}
\label{eq:19}
\fl \tilde{G}^{\alpha,\kappa}(x-y)=\sum_{n\in \mathbb{Z}} \frac{ C_n}{2\pi}
 e^{-i(\alpha+2n+1)
           [\pi\rho^0\lambda(x^1-y^1)+\arg(x-y)]} 
 |x-y|^{-\frac{\lambda^{-1}+(\alpha+2n+1)^2\lambda}{2}},
\end{eqnarray}
where $C_n$'s are some regularization parameters depending on the
details of UV limit of specific models. Notice that the left- and right-
handed fermions fall in the sectors with $n=-1$ and $n=0$,
respectively.  One may interpret the Green's function Eq.~\eref{eq:19}
as an anyon version in the Haldane's harmonic-fluid theory of
one-dimension quantum gas\cite{Haldane1981a,Haldane1981b}. Indeed, by
setting $\alpha=0$ or $\alpha=1$, it reduces to the well-known results
for fermions and bosons, respectively.

\section{Fractional statistics for a chiral anyon system}
\label{sec:nonm-stat-single-1}
The results given in Sec.~\ref{sec:chern-simons-theory} and
~\ref{sec:fract-braid-excl} show unambiguously that there is no direct
relation between braid and exclusion statistics with or without
interactions, if the one-dimensional matter field couples to the
statistical field gauge invariantly. However, if the system is anomalous
like chiral fermion, the fractional exclusion statistics can be induced
by braiding the free particles(chiral fermion) through the Chern-Simons
statistical transmutation, which is closely connected with our previous
study in two-dimensional systems\cite{Ye2015}, as we demonstrate in this
section.

We consider the simple case of chiral fermions. It has been shown long
time ago that one cannot couple gauge-invariantly the chiral fermions to
a gauge field $A_\mu$\cite{Alvarez1984}. A way out is to consider the
Dirac operator acting on the full mode-space of a (1+1)-dimensional,
two-component Dirac field and restrict the gauge field to its chiral
component
$(A_\mu \pm i\epsilon_{\mu\nu} A^\nu)/2\equiv A^{\pm}_{\mu}/2$. The
corresponding action has the following form:
\begin{eqnarray}
\label{eq:20}
S^{c\pm}_f[\psi,\bar{\psi}|A]
=\int d^2x
  \bar{\psi}i\gamma^{\mu}\bigg(\partial_{\mu}+i \frac{A^\pm_{\mu}}{2}
  \bigg)\psi -i\int d^2x A_0^{\pm}\rho^0_{\pm},
\end{eqnarray}
where $\rho^0_{\pm}$ is the density of right- and left- handed fermion,
respectively. Following Ref. \cite{Jackiw1985}, we integrate over the
chiral fermion field and obtain the effective action of the gauge field
\begin{eqnarray}
\label{eq:21}
  S_{eff}^{c\pm}[A_{\mu}]=
S_{eff}[\frac{A^{\pm}}{2}] +
  \frac{c}{8\pi}\int d^2x A_{\mu}A^{\mu}  -i\int d^2x A_0^{\pm}\rho^0_{\pm},
\end{eqnarray}
where $S_{eff}[A]$ is given in Eq.~\eref{eq:12} and a local quadratic
term of $A_{\mu}$ is added, reflecting a finite renormalization ambiguity
due to the lack of gauge invariance with the coefficient $c$, a priori an
arbitrary real constant. Here we take the "minimal choice" $c=1$.

The effective actions $S^{c\pm}_{eff}[A_{\mu}]$ for gauged chiral
fermions are anomalous: performing a gauge transformation,
$A^\mu \rightarrow A^\mu + \partial^\mu \Lambda$, we have
\begin{eqnarray}
\label{eq:22}
  S_{eff}^{c \pm}(A_\mu) \rightarrow S_{eff}^{c \pm}(A_\mu) \pm
  \frac{i}{4 \pi} \int d^2 x \Lambda \epsilon_{\mu\nu} \partial^\mu  A^\nu(x). 
\end{eqnarray} 
A remedy for such an inconsistency is to take the chiral fermion system
as the boundary of a bulk system with Hall conductance $\pm 1/(2\pi)$.
The bulk effective action then reads
\begin{eqnarray}
\label{eq:23}
  S_{bulk}^\pm[A] 
  =\pm\frac{i}{4\pi} \int d^3x \theta(x^2)
     \epsilon^{\mu\nu\lambda}A_{\mu}\partial_{\nu}A_{\lambda} 
   -i\int d^3x \theta(x^2)\rho^{0}_B A_0,
\end{eqnarray}
where $\theta(x^2)$ is the Heaviside step function and $\rho^0_B$ is the
expectation value of the fermion density in the bulk w.r.t. the
\emph{noninteracting} vacuum. The bulk action is also gauge variant due
to the existence of the boundary. In fact under the gauge
transformation, one finds
\begin{eqnarray}
\label{eq:24}
S_{bulk}^{\pm}[A] \rightarrow S_{bulk}^{\pm}[A] \mp \frac{i}{4\pi} \int
  d^2 x \Lambda \epsilon_{\mu\nu} \partial^\mu  A^\nu(x). 
\end{eqnarray} 
Comparing Eq.~\eref{eq:24} and Eq.~\eref{eq:22}, we observe the
anomaly of the bulk effective action restricted in the region $x^2>0$
cancel that of the matter field on its boundary. The whole action is
then gauge invariant.

Next, we couple the matter field to the statistical gauge field with
support in the whole space. After integrating out $A_0$ we obtain the
following constraint:
\begin{eqnarray}
\label{eq:25}
\fl i\delta(x^2)[j^{\pm}(x)+i\rho^{0}_{\pm}]-\rho_B^0\theta(x^2) 
=-\frac{F_{12}}{2\pi\alpha} \mp \frac{F_{12}}{2\pi}\theta(x^2) \pm
  \frac{1}{4\pi}\delta(x^2) A_1(x,x^2) 
\end{eqnarray}
where $j^{\pm}(x)\equiv[j^0(x)\pm ij^1(x)]/2$ is the chiral current on
the edge.

Before proceeding to the analysis of the statistics of chiral fermion,
we first examine the fractional statistics of the bulk fermion.  In the
bulk region $x^2>0$, the constraint reduces to
$F_{12}=2\pi\alpha\rho^0_B/(1\pm\alpha)$. Note that the particle density
$\rho_B(\alpha)$ in the ground state depends on $\alpha$ and the flux
density reads $F_{12}=2\pi\alpha\rho_B(\alpha)$ if the bulk system is
incompressible. Therefore, we find the relation between $\rho_B(\alpha)$
and the bare particle density $\rho^0_B$ for the bulk
\begin{eqnarray}
\label{eq:26}
\rho_B(\alpha)= \rho_B^0/(1\pm \alpha). 
\end{eqnarray}
which indicates that the Haldane's statistical interaction $g$ is
$\mp\alpha$.  The relation between $g$ and $\alpha$ for a Hall insulator
with a general Hall conductance $\sigma_h$ has been given in a different
way in our previous study\cite{Ye2015}. Inspired by the bulk result, one
may expect that the edge mode may also exhibit non-trivial fractional
\emph{exclusion} statistics.  Let us prove it by focusing only on the
right-handed anyon.

The gauge invariant Green's function for right-handed fermion in the
presence of Chern-Simons term reads
\begin{eqnarray}
\label{eq:27}
  \fl G^{\alpha}_{++}(x-y)
    = \frac{ \int \mathcal{D}\psi\mathcal{D}\bar{\psi}\mathcal{D}A
    \tilde{\psi}_R(x)\tilde{\psi}^{\dagger}_R(y)e^{ik_f(x^1-y^1)}
    e^{S_f^{c+}[\psi,\bar{\psi}|A]+S^+_{bulk}[A]+ S_{c.s.}[A]}}{\int
    \mathcal{D}\psi \mathcal{D}\bar{\psi}\mathcal{D}Ae^{S_f^{c+}[\psi,\bar{\psi}|A]+S^+_{bulk}[A]+ S_{c.s.}[A]}}.
\end{eqnarray}
One can notice that there is only one Fermi wavevector
$k_f=2\pi\rho_0^+$ for the chiral fermions, unlike for the Dirac
fermions where two Fermi points exist.  In Eq.~\eref{eq:27}, the
left-handed fermion is also present, however it is only an auxilliary
free field to make the fermion measure well defined\cite{Jackiw1985}. In
fact, the final result of $G_{++}^{\alpha}(x-y)$ defined in
Eq.~\eref{eq:27} is completely anti-analytic without interference
from the left-handed fermions. In order to implement the
$\alpha$-braiding, we put the gauge string of the anyon field
$\tilde{\psi}_a(x)$ outside the bulk to avoid the entanglement with the
bulk fermions.

To calculate the Green's function, we follow the same procedure given in
previous two sections. First, we solve the contraint Eq.~\eref{eq:25}
in the gauge $A_2=0$, leading to the following solution of the
statistical gauge field for the right-handed anyon
\begin{eqnarray}
\label{eq:28}
  A_1(x,x^2)=&i\pi\alpha \sgn(x^2)[j^{+}(x)+i\rho^0_+] - \theta(x^2) x^2
           2\pi\alpha\rho_B(\alpha), 
\end{eqnarray}
where we take $\theta(0)=0$. Substituting Eq.~\eref{eq:28} into
Eq.~\eref{eq:27} and using the Schwinger formula Eq.~\eref{eq:10}, the
Green's function can be written as
\begin{eqnarray}
\label{eq:29}
&G^{\alpha}_{++}(x-y) \nonumber\\
=&\frac{ \int \mathcal{D}\psi\mathcal{D}\bar{\psi}
\psi_R(x)\psi^{\dagger}_R(y)e^{i2\pi\rho_{+}^0(x^1-y^1)(1+\frac{\alpha}{2})}
    e^{\int d^2z\bar{\psi}i\gamma^{\mu}(\partial_{\mu}+i\mathcal{A}_{\mu})\psi}}{\int
  \mathcal{D}\psi
  \mathcal{D}\bar{\psi}e^{\int
   d^2z\bar{\psi}i\gamma^{\mu}\partial_{\mu}\psi}} \nonumber\\
=& \frac{\delta_{ab}}{2\pi} \frac{1}{(x^0-y^0)-i(x^1-y^1)}
e^{i\Theta_{x,y}^1[\mathcal{A}]}e^{S_{eff}[\mathcal{A}]},
\end{eqnarray}
where the auxilliary field $\mathcal{A}_{\mu}$ defined by
$\mathcal{A}_0(z) = \frac{\pi\alpha}{2}\partial_0Q_{x,y}(z)$ and
$\mathcal{A}_1(z) = i \frac{\pi\alpha}{2}\partial_0Q_{x,y}(z)$.  The
functionals $\Theta_{x,y}^1[\mathcal{A}]$ and $S_{eff}[\mathcal{A}]$ are
given in Eq.~\eref{eq:11} and Eq.~\eref{eq:12}, respectively, which
can be calculated straightforwardly. Thus we obtain the Green's function
for the right-handed anyon:
\begin{eqnarray}
\label{eq:30}
  G^{\alpha}_{++}(x-y)
  =&\frac{1}{2\pi} 
     \frac{e^{i2\pi\rho_+^0(1+\alpha/2)(x^1-y^1)}}{[x^0-y^0-i(x^1-y^1)]^{(1+
     \alpha/2)^2}}.
\end{eqnarray}

This result shows that the Fermi momentum $k_{f}$ is shifted from
$2\pi\rho_0^+$ to $2\pi\rho_0^+(1+\alpha/2)$ by the Chern-Simons
coupling. Since the chiral fermion has only one Fermi wavevector $k_f$,
its Fermi area is in fact linearly dependent on $k_f$. Therefore, the
shift of $k_f$ actually implies a non-trivial \emph{exclusion}
statistics. Notice that the Green's function has a power-law dependence
on the anti-holomorphic coordinates $x^0-ix^1$ with a fractional
exponent $(1+\alpha/2)^2$ which also reflects the braiding effect. The
present system is in fact a chiral ``Luttinger'' anyon liquid.

There is also an essential difference between the chiral and Dirac anyon
gases, namely, the braid phase factor arising from an oriented exchange
of the coordinates $x$ and $y$ for the chiral anyon depends
quadratically on the Chern-Simons coupling parameters $\alpha$. This
spoils the periodicity of the braid statistics on $\alpha$ seen in the
Dirac anyon system, as discussed in the previous sections. This can be
attributed to the lack of the backscattering channel for chiral
anyons. Indeed, the existence of two Fermi points is crucial for the
Dirac fermions to form different sectors with multiple particle-hole
excitations carrying a current of $2nk_f$($n\in \mathbb{Z})$, and the
shift of $\alpha$ by $2n$ simply transfers one sector to the other with
the same braid statistics.  Furthermore, in a conventional Luttinger
liquid, a Galileian boost can be used to pump one particle from one
Fermi point to the other, and to increase the total current by
$2k_f$\cite{WuYu1995} without changing the total particle
numbers. However, for a single branch of chiral fermion with one Fermi
point, such a boost not only increases the total current but also
changes the total particle number as it is connected with a bulk
particle reservior. Thus, with the particle number constraint, it is
simply not allowed to make such a boost, and one can not expect the
existence of different sectors.

We would like to stress that the present chiral ``Luttinger'' anyon
liquid is induced by pure Chern-Simons coupling, unlike the edge mode in
the fractional quantum Hall system where the chiral Luttinger liquid is
due to the interaction of electrons in the lowest Landau level. 

\emph{Remark}: One may also implement the braid and exclusion statistics
in an analogous way to fractional quantum Hall systems, and it turns out
$\alpha=\lambda^{-1}$. Therefore, the relation between $\alpha$ and $g$ in
the anomalous systems is not universal, depending on how the braid
statistics is implemented. Here, we just sketch such a different
implementation, and more details with a general physical interpretation
will be discussed elsewhere. With the same methods given in
Sec.~\ref{sec:fract-braid-excl}, one can easily prove that a Luttinger
fermion with Haldane parameter $\lambda$ coupled chirally to a gauge
field has an effective action given by $\lambda S_{eff}^{c
  \pm}(A_\mu)$. Then one can cancel its anomaly by adding a Chern-Simons
action for the gauge field in the half-space-time with a braid parameter
matching with $\lambda$. One can then construct a chiral field by
attaching to, e.g., $\psi_R(x)$ a phase string:
$\psi_{R}(x) \exp[- i (\lambda^{-1}-1) \pi \int_{x^1}^{\infty} d y^1
j^0(x^0,y^1)]$.  One finds that this field is indeed chiral with both braid
and fractional exclusion statistics.

\section{Mutual statistics of multiple species of chiral anyons}
\label{sec:mutu-stat-mult}
In this section we consider the mutual statistics among multiple chiral
anyon species. Multiple chiral edge modes may exist in the integer
quantum Hall insulators with Hall conductance $\sigma_h\ge 2$.  It may
also occur in the fractional quantum Hall systems in the hierarchical
theory, where the quantum state at some filling is not of the simple
Laughlin type, leading to many branches of edge
excitations\cite{Haldane1983,Frohlich1991,Wen1992}.

We consider the following action of right-handed fermions with $N_f$
flavors, each of which couples with the same statistical gauge field
with different statistical charges $q_a$,
\begin{eqnarray}
\label{eq:31}
\fl S_{N_f}^{c+}[\{\psi_{a},\bar{\psi}_{a}\}|A] 
=\sum_{a=1}^{N_f} \int d^2x \left[\bar{\psi}_ai\gamma^{\mu} 
\left( \partial_{\mu} + i q_a \frac{A^+_{\mu}}{2} \right)\psi_a 
-iq_aA_0^+\rho^0_{a+}\right].
\end{eqnarray}
As explained in Sec.~\ref{sec:nonm-stat-single-1}, for each right-handed
fermion, we need to add the fermion with opposite chirality which,
however, does not couple to the gauge field and serves as an auxilliary
field to give a well-defined fermionic integration measure.  The
anomalous effective action of statistical gauge field is simply
$\nu_t S_{eff}^{c+}[A_{\mu}]$ with
$\nu_t\equiv \sum_{a=1}^{N_f}q_a^2$(here we temporarily ignore the
density term purposely). To cancel the gauge anomaly of the chiral edge
modes and make the whole theory gauge invariant, we need a bulk
Chern-Simons term in the upper half plane with $x^2>0$
\begin{eqnarray}
\label{eq:32}
\fl S^+_{bulk}[A] = \frac{i\nu_t}{4\pi}\int d^3x \theta(x^2)
  \epsilon^{\mu\nu\lambda} A_{\mu}\partial_{\nu}A_{\lambda} 
-i \sum_{a=1}^{N_f}\int d^3x\theta(x^2) q_a\rho_{B,a}^0 A_0,
\end{eqnarray}
where $\rho_{B,a}^0$ is the bulk density of the $a$-flavor fermion
w.r.t.  the non-interacting vacuum. 

Next, we add the Chern-Simons action of Eq.~\eref{eq:2} in the whole
space to implement the braid statistics. Integrating out $A_0$ one finds
a simple extension of Eq.~\eref{eq:25},
\begin{eqnarray}
\label{eq:33}
&i\delta(x^2)\sum_{a=1}^{N_f}q_a[j_a^{+}(x)+i\rho^{0}_{a,+}]
-\sum_{a=1}^N\theta(x^2)q_{a}\rho_{B,a}^0
\nonumber\\
=&\frac{F_{12}}{2\pi\alpha} - \frac{\nu_tF_{12}}{2\pi}\theta(x^2) +
  \frac{\nu_t}{4\pi}\delta(x^2) A_1(x,x^2),
\end{eqnarray}
where $j^{+}_a(x)\equiv (j^0_a+ij^1_a)/2$ is the particle density of
right-handed fermions and the charge density is $q_aj^+_a$. For the bulk
region, this constraint has a simple form
$F_{12}=2\pi\alpha(1+\alpha\nu_t)\sum_{a=1}^{N_f}q_a\rho_{B,a}^0$. As we
proved in Ref.\cite{Ye2015}, the bulk anyon also obeys the mutual
fractional exclusion statistics induced by braiding particles, and the
particle density is shifted to a $\alpha$-dependent value
$\rho_{B,a}(\alpha)$ for each flavor. The corresponding total flux
density is then $2\pi\alpha\sum_{a=1}^{N_f}q_a\rho_{B,a}(\alpha)$,
therefore we obtain a relation between $\rho_{B,a}^0$ and
$\rho_{B,a}(\alpha)$ as following
\begin{eqnarray}
\label{eq:34}
\sum_{a=1}^{N_f}q_a\rho^0_{B,a} =
  (1-\alpha\nu_t)\sum_{a=1}^{N_f}q_a\rho_{B,a}(\alpha).
\end{eqnarray}
This result can also be derived using our previous results on the mutual
statistics in two-dimensional Hall insulator consisting of multiple
species of anyons, where we proved the parameters $g_{ab}$ of mutual
exclusion satisfy $g_{ab} =2\pi\sigma_{h,a} \alpha q_aq_b$ for all
flavors in the presence of Chern-Simons coupling(see the appendix of
Ref. \cite{Ye2015}).

We now turn to the mutual statistics of one-dimensional chiral
anyons. The solution of Eq.~\eref{eq:33} is similar to that of
Eq.~\eref{eq:28}, and in the gauge $A_2=0$ we have:
\begin{eqnarray}
\label{eq:35}
  A_1(x,x^2)=&i\pi\alpha \sgn(x^2)\sum_{a=1}^{N_f}q_a[j^{+}_a(x)+\rho^0_{a+}], 
\end{eqnarray}
where we omit the bulk density term, since it is not necessary for the
discussion on the mutual statistics of the edge anyons. Following the
same procedure given for the single chiral fermion, it is
straightforward to derive the Green's function of the chiral fermions
\begin{eqnarray}
\label{eq:36}
& G^{\alpha,+}_{ab}(x-y)  \nonumber\\
 =& \frac{\delta_{ab}}{(Z_{0f})^{N_f}}
\int \mathcal{D}\psi\mathcal{D}\bar{\psi}
e^{i2\pi\rho_a^+(x^{1}-y^1)+i\pi\alpha q_a(x^1-y^1)\sum_{c=1}^{N_{f}}q_{c}\rho^0_{c+}} \nonumber\\
 &\times
\psi_{aR}(x)\psi^{\dagger}_{aR}(y)e^{\sum_{c=1}^{N_f}\int d^{2}z\bar{\psi}_ci\gamma^{\mu}(\partial_{\mu}+iq_c\mathcal{A}_{\mu})\psi_c}
\nonumber\\
 =& \frac{\delta_{ab}}{2\pi}
 \frac{e^{i2\pi
(\rho_a^++ \frac{\alpha q_a}{2} \sum_{c=1}^{N_f}q_{c}\rho_{c+}^0
    )(x^1-y^1)}}
{[x^0-y^0-i(x^1-y^1)]^{1+\alpha q_a^2+ \frac{q_a^2}{4}\sum_{c=1}^{N_f}q^2_c\alpha^2}},
\end{eqnarray}
where $\mathcal{A}_0= \pi\alpha q_a\partial_0Q_{x,y}(z)/2$ and
$\mathcal{A}_1=i\pi\alpha q_a\partial_0Q_{x,y}(z)/2$ coming from the
the solution of the constraint Eq.~\eref{eq:35}.  

There is no correlation between anyons with different flavors. However
the Green's function gets modifications from other flavors of chiral
fermions: (1) The anomalous exponent of the Green's function
$G^{\alpha,+}_{aa}(x-y)$ for flavor $a$ receives the contributions from
other chiral fermions, which is $\sum_{c\ne a}(q_aq_c\alpha/2)^2$.  This
reflects the interaction induced by the flux binding between different
flavors of chiral fermions. (2) The Fermi momentum $k_f^a$, which
reflects the occupation status of flavor $a$ particle in the momentum
space, is also modified by other chiral fermions.  The change of the
Fermi momentum $k_f^a$ w.r.t. $\rho_b^+$ can be calculated
straightforwardly:
\begin{eqnarray}
\label{eq:37}
\frac{\partial k_f^a}{\partial \rho_b^+} = 2\pi\delta_{ab} +
  \pi\alpha q_aq_b. 
\end{eqnarray}
where the second term unambiguously shows the mutual exclusion
statistics consistent with our previous results for the two-dimensional
Hall systems\cite{Ye2015}. The present mutual exclusion statistics is
indeed induced only by the mutual exchange statistics, as we do not add
any interactions.

\section{T-duality and Bosonization}
\label{sec:bosonization}
In this section we review some standard formulas of one-dimensional
bosonization in the Euclidean path-integral formalism following
Ref.\cite{Frohlich1988b}, and we apply them to the one-dimensional
"Luttinger" anyons given in Sec.~\ref{sec:fract-braid-excl} and chiral
anyons in Sec.~\ref{sec:nonm-stat-single-1}. Then, the results given in
the previous sections can be reproduced.

We introduce the zero-mass Gaussian measure in euclidean 1+1 spacetime
with mean zero and covariance $(4\pi \lambda)(-\Delta)^{-1}$, where
$\lambda>0$. This measure is written formally as
\begin{eqnarray}
\label{eq:38}
\frac{1}{Z} \int {\mathcal D} \phi  e^{\frac{1}{8 \pi\lambda} \int d^2 x \partial_\mu \phi(x) \partial^\mu \phi (x)},
\end{eqnarray}
assuming the scalar field $\phi$ to vanish at infinity and $Z$ is the
partition function of free boson. Denoting by
$\langle \cdot \rangle_\lambda$ the corresponding expectation value, the
Gaussian measure can be more rigorously defined by:
\begin{eqnarray}
  \langle e^{i\int d^2 x  \phi(x) f(x)} \rangle_\lambda = e^{ 2 \pi \lambda \int d^2 x d^2 y f(x) \Delta^{-1} (x,y) f(y)},
\end{eqnarray}
if $f$ is a test function whose Fourier transform vanishes at the
origin, and $\langle e^{i\int d^2 x \phi(x) f(x)} \rangle_{\lambda} = 0$
if $f$ is real with non-vanishing Fourier transform at the origin.  We
now introduce the two main composite fields we will use in the theory
with expectation value $\langle \cdot \rangle_\lambda$. The first is the
vertex, which is just an imaginary exponential of the field $\phi$
normal ordered, formally defined by
$ :e^{i \beta \phi(x)}:= e^{i \beta \phi(x)}(2 \pi)^{\beta^2}
e^{-2 \pi\lambda \beta^2 \Delta^{-1} (x,x) }$, for
$\beta \in  \mathbb{R}$. The second one is the disorder field. Let us consider
the vector potential $V_\mu^x(y)$ of a magnetic vortex of charge 1 at
the point $x$. It satisfies
$\epsilon^{\mu\nu}\partial_\mu V_{\nu}^{x}(y)=\delta(x-y)$ and locally on
$\mathbb{R}^2\backslash \{x\}$ $\sim \mathbb{C}\backslash \{x\}$ can be written
as $V_{\mu}^{x}(y)=\partial_\mu \arg(x-y)/ (2 \pi)$. The expectation value
of a product of $N$ disorder fields
$D(x^j, \zeta^j), \zeta^j \in \mathbb{R}$, $j=1...N$ is given, up to a UV
renormalization, by
\begin{eqnarray}
\left\langle \prod_{j=1}^N D(x^j, \zeta^j)\right\rangle_\lambda = \frac{\int {\mathcal D} \phi  e^{-\frac{1}{8 \pi\lambda} \int d^2 z (\partial_\mu \phi + \sum_{j=1}^N \zeta^j V^{x^j} _\mu)^2(z)}}{\int {\mathcal D} \phi  e^{-\frac{1}{8 \pi\lambda} \int d^2 z (\partial^\mu \phi)^2(z)}}. 
\end{eqnarray}
If both vertex and disorder fields are present one gets
\begin{eqnarray}
\label{eq:39}
\fl &  \left\langle \prod_{j=1}^N D(x^j,\zeta^j) \prod_{\ell=1}^{M}:e^{i \beta^\ell
  \phi(y^\ell)}:\right\rangle_\lambda \nonumber\\
\fl  =& \frac{\int {\mathcal D} \phi  e^{-\frac{1}{8 \pi\lambda} \int d^2 z (\partial_\mu \phi + \sum_{j=1}^N \zeta^j V^{x^j} _\mu)^2(z)}\prod_{\ell=1}^M e^{i \beta^\ell [\phi(y^\ell)+\sum_{j=1}^N \zeta^j \arg(x^j-y^\ell)]}}{\int {\mathcal D} \phi  e^{-\frac{1}{8 \pi\lambda} \int d^2 z (\partial^\mu \phi)^2(z)}} 
\end{eqnarray}
Performing the integration over $\phi$ one evaluates the version of
Eq.~\eref{eq:39} relevant for the bosonization of anyons' two-point
function obtaining:
\begin{eqnarray}
\label{eq:40}
\fl \langle D(x, \zeta) :e^{i \beta \phi(x)}: D(y, -\zeta) :e^{- i \beta
  \phi(y)}:\rangle_\lambda 
=
e^{-( \frac{\zeta^2}{2\lambda}+2\lambda \beta^2)\ln\vert x-y \vert}e^{-i 2\zeta \beta \arg(x-y)}
\end{eqnarray}

Next we show the bosonization as a special version of T-duality(see
e.g. Ref.\cite{Alvarez1995}) in one dimension, as has been realized in
Ref.\cite{Burgess1994a} and independently in
Refs.\cite{Frohlich1995,Marchetti1995a}. The basic idea underlying
T-duality is the following: Consider a quantum field theory expressed in
euclidean formalism in terms of charged fields $\chi, \chi^{*}$ whose
action $S(\chi, \chi^{*})$ is invariant under an abelian (e.g.  U(1))
global gauge transformation
\begin{eqnarray*}
\chi(x) \rightarrow e^{i \xi}\chi(x), \chi^{*}(x) \rightarrow e^{-i \xi}\chi^{*}(x).
\end{eqnarray*}
Then, we promote the global gauge invariance to a local gauge invariance
by introducing a minimal coupling between $\chi, \chi^{*}$ and a
U(1)-gauge field $C_\mu$. We then integrate over $C_\mu$ with the
zero-field constraint $\epsilon^{\mu\nu}\partial_\mu C_\nu = 0$, so that
the original theory is recovered. The Lagrange-multiplier field
enforcing the constraint for $C_{\mu}$ is the boson field of the
corresponding T-dual theory. Bosonization is just T-duality in case when
$\chi$ is the fermion field $\psi$. Let us show this procedure for the
partition function of "Luttinger" anyons with formal calculation. Their
partition function Eq.~\eref{eq:15} can be rewritten as:
\begin{eqnarray}
Z_T=&\int {\mathcal D}A e^{{S}_{cs}[A]} \delta (\partial^{\nu} A_\nu)
\int{\mathcal D} C \delta (\epsilon^{\mu\nu}\partial_\mu
  C_\nu)\delta(\partial^\nu C_\nu) 
\nonumber\\
&\int{\mathcal D}B e^{\int d^2 z \frac{1}{2 \kappa \pi} B_\mu B^{\mu} }
\int{\mathcal D} \psi {\mathcal D} \bar\psi e^{\mathcal{S}_f[\psi, \bar\psi \vert A+B+C]},
\end{eqnarray}
where $\delta(\partial^\nu C_\nu)$ and $\delta(\partial^{\nu}A_{\nu})$
are just possible gauge-fixing for $C_\mu$ and $A_{\mu}$, respectively.
For Dirac fermion, $S_f$ is given by Eq.~\eref{eq:1}, and one can
integrate the fermion fields and H.S. field firstly leading to
\begin{eqnarray*}
Z_T=&\int {\mathcal D}A e^{{S}_{cs}[A]} \delta (\partial^{\nu} A_\nu)
\int{\mathcal D} C \delta (\epsilon^{\mu\nu}\partial_\mu
  C_\nu)\delta(\partial^\nu C_\nu)  \nonumber\\
&\times
e^{\frac{1}{1+\kappa}\{S_{eff}[A+C]-i\int d^2z (A_{0}+C_{0})\rho^0\}}
\end{eqnarray*}
with $S_{eff}[A]$ given in Eq.~\eref{eq:12}. We then represent the
gauge-invariant constraint on $C$ as:
\begin{eqnarray}
\delta (\epsilon^{\mu\nu}\partial_\mu C_\nu)=\int {\mathcal D} \phi  e^{-i \int d^2z \frac{1}{2 \pi}\phi(z)\epsilon^{\mu\nu}\partial_\mu C_\nu(z)},
\end{eqnarray} 
where the Lagrange-multiplier $\phi$ is a real scalar field and the
factor $2 \pi$ has been introduced for later convenience. Changing
variable from $C_\mu+A_\mu \rightarrow C_\mu$ and subsequently
integrating over $C$, one obtains
\begin{eqnarray}
\label{eq:41}
Z_T=\int {\mathcal D}A e^{{S}_{cs}(A)} \delta (\partial^\nu A_\nu)
    \int{\mathcal D}\phi e^{\int d^2 z \mathcal{L}_B[\phi,A]}, \nonumber\\
\mathcal{L}_B[\phi,A]=\frac{1 +\kappa}{8 \pi}
      \partial_\mu \phi\partial^{\mu}\phi+\frac{\rho^0}{2}
                        \epsilon_{0\nu}
                        \partial^{\nu}
                        \phi + \frac{i}{2 \pi} \epsilon^{\mu\nu}\partial_\mu \phi A_\nu
\end{eqnarray}
For $A=0$, one recognizes the bosonized Luttinger liquid action with
$\lambda= 1/(1+\kappa)$, plus a density term. To keep the calculation
well defined, we provide the particle density with a support in a finite
system of length $L$, and eventually take the thermodynamic limit. Then,
one can shift
$\phi(x)\rightarrow
\phi(x)-2\pi\lambda\int_{-\infty}^{x^1}dz^1\rho^0(z^1)$, and the boson
Lagrangian acquires a standard Gaussian form given in Eq.~\eref{eq:38}
\begin{eqnarray*}
\mathcal{L}_B[\phi,A]\rightarrow 
\frac{1}{8\pi\lambda}\partial_{\mu}\phi\partial^{\mu}\phi - \frac{\pi\lambda(\rho^0)^2}{2},
\end{eqnarray*}
where the additional quadratic density term reminds us of the charge
excitation given in Haldane's Luttinger liquid
theory\cite{Haldane1983}. When calculating the correlation function of
vertex operators, the previous shift contributes an additional phase
factor given below,
\begin{eqnarray}
\label{eq:42}
\fl \frac{\int \mathcal{D}\phi e^{i \beta[\phi(x)-\phi(y)]}
  e^{\int d^2z \mathcal{L}[\phi,0]} }{\int \mathcal{D}\phi 
  e^{\int d^2z \mathcal{L}[\phi,0]} }
=e^{-i2\pi\lambda\beta\rho^0(x^1-y^1)}
\langle :e^{i\beta\phi(x)}::e^{-i\beta\phi(y)}:\rangle_{\lambda}
\end{eqnarray}
which is useful to identify the exclusion statistics via the change of
``Fermi'' area as shown in previous sections, where this
$\rho^{0}$-related phase factor is calculated alternatively in the
fermion formalism.

In a similar way, one can
obtain the bosonized anyon two-point function with the identifications:
\begin{eqnarray}
\label{eq:43}
&\psi_a(x) \rightarrow D(x,1):e^{i \frac{[(-1)^a + \alpha] }{2} \phi(x)}:, \nonumber\\
&\psi^{\dagger}_a(x) \rightarrow D(x,-1):e^{-i \frac{[(-1)^a + \alpha] }{2} \phi(x)}:.
\end{eqnarray}

In the following, we sketch how to derive via duality the bosonization
formulas Eq.~\eref{eq:43} in the simplest case with $\kappa=0$, and the
general case can be easily handled by inserting the $B$ fields following
the procedure outlined in Sec.~\ref{sec:fract-braid-excl}(for more
details see Ref.\cite{Frohlich1995}).  First, for $\alpha=0$, the
Green's function of non-interacting fermion fields can be written in terms
of auxilliary field $C_{\mu}$ and boson field $\phi$,
\begin{eqnarray}
\label{eq:44}
& \left\langle {\psi}_a(x)\psi^{\dagger}_a(y)\right\rangle \nonumber\\
 =&
 \frac{\int
  {\mathcal D}\phi {\mathcal D} C \delta(\partial^\nu C_\nu)
e^{S_{eff}[C]-i \int
  d^2z \frac{1}{2 \pi}\phi\epsilon_{\mu\nu}\partial_\mu
  C_\nu}\psi_a(x)\psi^{\dagger}_b(y)
  }{\int {\mathcal
  D}\phi {\mathcal D} C \delta(\partial^\nu C_\nu) e^{S_{eff}[C]-\int
  d^2z \frac{i}{2 \pi}\phi\epsilon^{\mu\nu}\partial_\mu
  C_\nu}} \nonumber\\
 =&
\frac{e^{-i(-1)^a\arg(x-y)}}{|x-y|} 
 \frac{\int
  {\mathcal D}\phi {\mathcal D} C \delta(\partial^\nu C_\nu)
  e^{i\Theta^a_{x,y}[C]+S_{eff}[C]- \int d^2z \frac{i}{2
  \pi}\phi\epsilon^{\mu\nu}\partial_\mu C_\nu}}{\int {\mathcal
  D}\phi {\mathcal D} C \delta(\partial^\nu C_\nu) e^{S_{eff}[C]- \int
  d^2z \frac{i}{2 \pi}\phi\epsilon^{\mu\nu}\partial_\mu
  C_\nu}} \nonumber\\
  =& e^{-i(-1)^a \arg(x-y) - \pi \Delta^{-1}(x-y)} \nonumber\\
 &\times\frac{\int {\mathcal
    D}\phi e^{-\int d^2 z \frac{1}{8 \pi} (\partial_\mu \phi)^2
    (z)}:e^{i \frac{(-1)^a}{2}\phi(x)}::e^{-i \frac{(-1)^a}{2}\phi(y)}:
    }{\int{\mathcal D}\phi e^{-\int d^2 z
    \frac{1}{8 \pi} (\partial_\mu \phi)^2 (z)}} \nonumber\\
 =&\langle 
D(x,1):e^{i \frac{(-1)^a}{2}\phi(x)}:
D(y,-1):e^{-i\frac{(-1)^a}{2}\phi(y)}: 
 \rangle,
\end{eqnarray}
where we used Schwinger's formula Eq.~\eref{eq:10} and we set
$\rho^0=0$ since it is irrelvant for the present purpose.  For the anyon
field, we need to attach the gauge string. Using Eq.~\eref{eq:4} and
$j^0(x)=- i\partial_1\phi/(2\pi)$ in the bosonization form, one can
easily justify Eq.~\eref{eq:43}.

The bosonization for the chiral fermion based on T-duality is
similar. Comparing the Lagrangian for the chiral fermion given in
Eq.~\eref{eq:20} with that for the Dirac fermion, we find the bosonized
form for chiral fermion which is obtained by replacing $A_{\mu}$ in
Eq.~\eref{eq:41} with $A^{\pm}_{\mu}/2$,
\begin{eqnarray}
\label{eq:45}
\mathcal{L}_B^{\pm}=\frac{1}{8 \pi}
      \partial_\mu \phi\partial^{\mu}\phi+\frac{\rho^0_{\pm}}{2} \epsilon_{0\nu}\partial_\nu
      \phi + \frac{i}{4 \pi} \epsilon^{\mu\nu}\partial_\mu \phi
  A_{\nu}^{\pm} + \frac{c}{4\pi}A_{\mu}A^{\mu},
\end{eqnarray}
where the local quadratic term of $A_{\mu}$ is added again due to the
finite renormalization ambiguity. The boson Lagrangian
$\mathcal{L}_B^{\pm}$ can reproduce the same effective action
$S_{eff}^{\pm}[A]$ as the Lagrangian of chiral fermion, though it is not
the minimal version. To get the minimal form of boson Lagrangian
$\mathcal{L}_B^{\pm}$, one can simply put the chiral constraint on the
fermion current, then it is easy to reproduce in this T-dual formalism
the chiral bosonization of Refs.\cite{Floreanini1987,Harada1990a}.

\section{Conclusions and Remarks}
\label{sec:conclusions-remarks}
As a summary, we adopt the Chern-Simons gauge theory with suitable
dimensional reduction to clarify the relation between the braid and
fractional exclusion statistics in one dimension. The same framework has
also been used in our previous study on the two-dimensional
case\cite{Ye2015}, thus, completing a systematic study on the two
aspects of the fractional statistics in low dimensions $d\le2$.

For Dirac fermions, the flux-binding does not necessarily induce the
nontrivial fractional exclusion statistics, which is also consistent
with the result given in Ref.\cite{Aglietti1996}.  Here we would like to
mention that, for the exactly solvable Calogero-Sutherland model, which
exhibits explicitly the fractional exclusion statistics as derived from
its energy spectrum, one can actually assign arbitrary braid statistics
to its manybody wavefunction without changing the energy spectrum due to
the impenetrable $x^{-2}$ interaction. Indeed, the fermion and boson
solutions of this model were given in Ref.\cite{Sutherland1971c}, and
another ``natural'' anyonic Jastrow-Laughlin type wavefunction together
with the corresponding correlation functions was constructed in
Refs.\cite{Ha1994b,Ha1995}. Therefore, in this model, the two statistics are not
necessarily connected.

For chiral fermions, however, binding flux to the fermion field can induce
a well-defined fractional exclusion statistics for both single and
multiple species of particles. Since the chiral fermion is simply a
boundary system for a Hall insulator, and this result is consistent with
our study on the two dimensional cases\cite{Ye2015} where we proved the
braid statistics together with Hall response can result in a nontrivial
exclusion statistics. Both the one and two dimensional results suggest
that the time-reversal breaking in the original system before coupling
to a Chern-Simons term is somehow necessary for connecting the braid and
fractional exclusion statistics for dimensions $d\le2$.

Since the fractional statistics are natually related to interactions, we
hope our study may shed light on the application of fractional
statistics to strongly correlated condensed matter system within the
framework of Chern-Simons gauge theroy. In particular, using the
tomographic decomposition\cite{Frohlich1997} one can analyze the
long-range behaviour(scaling limit) of two-dimensional fermionic systems
in terms of one-dimensional systems labelled by the rays of the
two-dimensional Fermi surface. As an application of the formalism
developed here, we are presently considering the two-dimensional $t-J$
model, relevant for the high Tc cuprates.

\section*{Acknowledgements} 
F.Y. is supported by National Nature Science Foundation of China
11374135 and JCYJ20160531190535310.  P.A.M. gratefully thanks
J. Fr\"{o}hlich for pointing out some relevant references and acknowledges
the partial support from the Ministero Istruzione Universit\'a Ricerca
(PRIN Project "Collective Quantum Phenomena: From Strongly-Correlated
Systems to Quantum Simulators").
\section*{References}

\begin{thebibliography}{10}
\expandafter\ifx\csname url\endcsname\relax
  \def\url#1{{\tt #1}}\fi
\expandafter\ifx\csname urlprefix\endcsname\relax\def\urlprefix{URL }\fi
\providecommand{\eprint}[2][]{\url{#2}}

\bibitem{Wu1984a}
Wu Y~S 1984 {\em Phys. Rev. Lett.\/} {\bf 52}(24) 2103--2106

\bibitem{Frohlich1988a}
Fr{\"o}hlich J and Marchetti P~A 1988 {\em Letters in Mathematical Physics\/}
  {\bf 16} 347--358

\bibitem{Leinaas1977}
Leinaas J~M and Myrheim J 1977 {\em Nuovo Cimento B\/} {\bf 37} 1--23

\bibitem{Goldin1981}
Goldin G~A, Menikoff R and Sharp D~H 1981 {\em Journal of Mathematical
  Physics\/} {\bf 22} 1664--1668

\bibitem{Wilczek1982}
Wilczek F 1982 {\em Phys. Rev. Lett.\/} {\bf 49}(14) 957--959

\bibitem{Streater1970}
Streater R and Wilde I 1970 {\em Nuclear Physics B\/} {\bf 24} 561 -- 575

\bibitem{Frohlich1976}
Fr{\"o}hlich J 1976 {\em Communications in Mathematical Physics\/} {\bf 47}
  269--310

\bibitem{Fradkin1980}
Fradkin E and Kadanoff L~P 1980 {\em Nuclear Physics B\/} {\bf 170} 1 -- 15

\bibitem{Frohlich1992a}
Fr{\"o}hlich J 1992 {\em Non-Perturbative Quantum Field Theory Mathematical
  Aspects and Applications, World Scientific Advanced Series in Mathematical
  Physics: Volume 15\/} (World Scientific)

\bibitem{Polyakov1988}
Polyakov A~M 1988 {\em Mod. Phys. Lett.\/} {\bf A3} 325

\bibitem{Haldane1991a}
Haldane F~D~M 1991 {\em Phys. Rev. Lett.\/} {\bf 67}(8) 937--940

\bibitem{Wu1994}
Wu Y~S 1994 {\em Phys. Rev. Lett.\/} {\bf 73}(7) 922--925

\bibitem{WuYu1995}
Wu Y~S and Yu Y 1995 {\em Phys. Rev. Lett.\/} {\bf 75}(5) 890

\bibitem{Lieb1963a}
Lieb E~H and Liniger W 1963 {\em Phys. Rev.\/} {\bf 130}(4) 1605--1616

\bibitem{Lieb1963b}
Lieb E~H 1963 {\em Phys. Rev.\/} {\bf 130}(4) 1616--1624

\bibitem{yang1969}
Yang C~N and Yang C~P 1969 {\em J. Math. Phys.\/} {\bf 10} 1115--1122

\bibitem{Calogero1969b}
Calogero F 1969 {\em J. Math. Phys.\/} {\bf 10} 2197--2200

\bibitem{Sutherland1971a}
Sutherland B 1971 {\em J. Math. Phys.\/} {\bf 12} 246--250

\bibitem{Sutherland1971b}
Sutherland B 1971 {\em J. Math. Phys.\/} {\bf 12} 251--256

\bibitem{Sutherland1971c}
Sutherland B 1971 {\em Phys. Rev. A\/} {\bf 4}(5) 2019--2021

\bibitem{Sutherland1972}
Sutherland B 1972 {\em Phys. Rev. A\/} {\bf 5}(3) 1372--1376

\bibitem{deVeigy1994}
Dasni\`eres~de Veigy A and Ouvry S 1994 {\em Phys. Rev. Lett.\/} {\bf 72}(5)
  600--603

\bibitem{Ye2015}
Ye F, Marchetti P~A, Su Z~B and Yu L 2015 {\em Phys. Rev. B\/} {\bf 92}(23)
  235151

\bibitem{Bernard1994anote}
Bernard D and Wu Y~S 1994 A note on statistical interactions and the
  thermodynamic bethe ansatz, in "new developments in integrable systems and
  long-range interaction models" {\em Nankai Lecture Notes on Mathematical
  Physics, (World Scientific)\/} (\textit{Preprint} \eprint{cond-mat/9404025})

\bibitem{Wu2001}
Wu Y~S, Yu Y and Yang H~X 2001 {\em Nucl. Phys. B\/} {\bf 604} 551 -- 579

\bibitem{Iguchi1998}
Iguchi K 1998 {\em Phys. Rev. Lett.\/} {\bf 80}(8) 1698--1701

\bibitem{Laughlin1988}
Laughlin R~B 1988 {\em Science\/} {\bf 242} 525--533

\bibitem{Frohlich1992}
Fr\"ohlich J and Marchetti P~A 1992 {\em Phys. Rev. B\/} {\bf 46}(10)
  6535--6552

\bibitem{Marchetti1998}
Marchetti P~A, Su Z~B and Yu L 1998 {\em Phys. Rev. B\/} {\bf 58}(9) 5808--5824

\bibitem{Marchetti2011a}
Marchetti P~A, Ye F, Su Z~B and Yu L 2011 {\em EPL (Europhysics Letters)\/}
  {\bf 93} 57008

\bibitem{Ha1994b}
Ha Z~N~C 1994 {\em Phys. Rev. Lett.\/} {\bf 73}(12) 1574--1577

\bibitem{Ha1995}
Ha Z 1995 {\em Nucl. Phys. B\/} {\bf 435} 604 -- 636

\bibitem{Marchetti1996}
Marchetti P~A, Su Z~B and Yu L 1996 {\em Nucl. Phys. B\/} {\bf 482} 731

\bibitem{Marchetti2010}
Marchetti P 2010 {\em Foundations of Physics\/} {\bf 40} 746--764

\bibitem{Schwinger1962}
Schwinger J 1962 {\em Phys. Rev.\/} {\bf 128}(5) 2425--2429

\bibitem{Gogolin1998}
Gogolin A~O, Nersesyan A~A and Tsvelik A~M 1998 {\em Bosonization and Strongly
  Correlated Systems\/} (Cambridge University Press)

\bibitem{Aglietti1996}
Aglietti U, Griguolo L, Jackiw R, Pi S~Y and Seminara D 1996 {\em Phys. Rev.
  Lett.\/} {\bf 77}(21) 4406--4409

\bibitem{Haldane1981a}
Haldane F~D~M 1981 {\em J. Phys. C: Solid State Physics\/} {\bf 14} 2585

\bibitem{Haldane1981b}
Haldane F~D~M 1981 {\em Phys. Rev. Lett.\/} {\bf 47}(25) 1840--1843

\bibitem{Alvarez1984}
Alvarez-Gaum\'{e} L and Ginsparg P 1984 {\em Nuclear Physics B\/} {\bf 243} 449
  -- 474

\bibitem{Jackiw1985}
Jackiw R and Rajaraman R 1985 {\em Phys. Rev. Lett.\/} {\bf 54}(12) 1219--1221

\bibitem{Haldane1983}
Haldane F~D~M 1983 {\em Phys. Rev. Lett.\/} {\bf 51}(7) 605--608

\bibitem{Frohlich1991}
Fr\"{o}hlich J and Zee A 1991 {\em Nuclear Physics B\/} {\bf 364} 517 -- 540

\bibitem{Wen1992}
Wen X~G 1992 {\em Int. J. Mod. Phys. B\/} {\bf 06} 1711--1762

\bibitem{Frohlich1988b}
Fr{\"o}hlich J and Marchetti P 1988 {\em Communications in Mathematical
  Physics\/} {\bf 116} 127--173

\bibitem{Alvarez1995}
Alvarez E, Alvarez-Gaum\'{e} L and Lozano Y 1995 {\em Nuclear Physics B -
  Proceedings Supplements\/} {\bf 41} 1

\bibitem{Burgess1994a}
Burgess C and Quevedo F 1994 {\em Nuclear Physics B\/} {\bf 421} 373 -- 387

\bibitem{Frohlich1995}
Fr\"{o}hlich J, G\"{o}tschmann R and Marchetti P~A 1995 {\em Journal of Physics
  A: Mathematical and General\/} {\bf 28} 1169

\bibitem{Marchetti1995a}
Marchetti P~A 1995 {Bosonization and duality in condensed matter systems} {\em
  {3rd Chia Meeting on Common Trends in Condensed Matter and High-energy
  Physics Chia Laguna, Sardinia, Italy}\/} (\textit{Preprint}
  \eprint{hep-th/9511100})

\bibitem{Floreanini1987}
Floreanini R and Jackiw R 1987 {\em Phys. Rev. Lett.\/} {\bf 59}(17) 1873--1876

\bibitem{Harada1990a}
Harada K 1990 {\em Phys. Rev. Lett.\/} {\bf 64}(2) 139--141

\bibitem{Frohlich1997}
Fr\"ohlich J and G\"otschmann R 1997 {\em Phys. Rev. B\/} {\bf 55}(11)
  6788--6815

\end{thebibliography}
\providecommand{\newblock}{}

\end{document}